\documentclass[12pt,a4paper]{article}

\usepackage{amsmath,amsfonts,amssymb,bm}

\usepackage{graphicx} % do wczytywania rysunkow

\usepackage[hang,normalsize,bf]{subfigure} % rysuje kilka rysunkow
\begin{document}

\begin{center}

{\Huge \bf
Unintegrated gluon distributions \\
and Higgs boson production \\ in proton-proton collisions
}

\vspace {0.6cm}

{\large M. {\L}uszczak $^{2}$ and A. Szczurek $^{1,2}$}

\vspace {0.2cm}

$^{1}$ {\em Institute of Nuclear Physics PAN\\
PL-31-342 Cracow, Poland\\}
$^{2}$ {\em University of Rzesz\'ow\\
PL-35-959 Rzesz\'ow, Poland\\}

\end{center}

\begin{abstract}
Inclusive cross sections for Higgs boson production in proton-proton
collisions are calculated in the formalism of unintegrated gluon
distributions (UGDF). Different UGDF from
the literature are used. Although they were constructed in order to
describe the HERA deep-inelastic scattering $F_2$ data, they lead to
surprisingly different results for Higgs boson production. We present both
two-dimensional invariant cross section as a function of Higgs boson
rapidity and transverse momentum, as well as corresponding projections
on rapidity or transverse momentum.
We quantify the differences between different UGD's by applying
different cuts on interrelations between transverse momentum of Higgs
and transverse momenta of both fusing gluons.
We focus on large rapidity region.
The interplay of the gluon-gluon fusion and weak-boson fusion in
rapidity and transverse momentum is discussed. We find that
above $p_T \sim$ 50-100 GeV the weak-gauge-boson fusion dominates
over gluon-gluon fusion.
\end{abstract}


%---------------------
\section{Introduction}
%---------------------

Recently unintegrated gluon (parton) distributions became a useful
and intuitive phenomenological language for applications to many
high-energy reactions (see e.g. \cite{smallx1,smallx2} and references
therein).
Mostly the HERA $F_2$ data was used to test or tune
different models of UGDF's. However, the structure function data are
not the best tool to verify UGDF, in particular its dependence on
gluon transverse momentum, because it enters to the $\gamma^* p$ total
cross section in an integrated way. 
UGDF's have been used recently
to describe jet correlations \cite{SNSS01}, correlations in heavy quark
photoproduction \cite{LS04}, total cross section for Higgs
production \cite{jung04},
inclusive spectra of pions in proton-proton collisions \cite{szczurek03}
or even nucleus-nucleus collisions \cite{KL01}.
It is rather obvious that differential cross sections seem
a much better tool than total or integrated cross sections
to verify UGDF's.

Many unintegrated gluon distributions in the literature
are ad hoc parametrizations of different sets of experimental data
rather than derived from QCD.
An example of a more systematic approach, making use of familiar
collinear distributions can be found in Ref.\cite{KMR01}.
Recently Kwieci\'nski and collaborators
\cite{CCFM_b1,CCFM_b2,GKB03} have shown how to solve so-called
CCFM equations by introducing unintegrated parton distributions in
the space conjugated to the transverse momenta \cite{CCFM_b1}.
We present results for inclusive Higgs production based on unintegrated
gluon distributions obtained by solving a set of coupled equations
\cite{GKB03}.
Recently these parton distributions were tested
for inclusive gauge boson production in proton-antiproton
collisions \cite{KS04} and for charm-anticharm correlations
in photoproduction \cite{LS04}. 

While in the gauge boson production one tests mainly quark and
antiquark (unintegrated) distributions at scales 
$\mu^2 \sim M_W^2, M_Z^2$, in the charm-quark photoproduction
one tests mainly gluon distributions at scales $\mu^2 \sim m_c^2$. 
The nonperturbative aspect of UPDF's can be tested for soft
pion production in proton-proton collisions \cite{CS05}. 

Different ideas based on perturbative and nonperturbative QCD
have been used in the literature to obtain the unintegrated gluon
distributions in the small-x region. Since almost all of them were
constructed to describe the HERA data it is necessary to test
these distributions in other high-energy processes in order to
verify the underlying concepts and/or approximations applied.
It is the aim of this paper to show predictions of these quite
different UGDF's for Higgs production at LHC at CERN
although we are aware of the fact that a real test against
future experimental data may be extremely difficult.
We compare and analyze two-dimensional distributions for inclusive
Higgs production in rapidity and transverse momentum $(y,p_t)$ to
study a potential for such an analysis in the future. We focus not
only on midrapidities but also try to understand a potential for
studying UGDF's in more forward or backward rapidity regions.
The results of the gluon-gluon fusion are compared with other
mechanisms of Higgs boson production such as WW fusion for instance.

%------------------
\section{Formalism}
%------------------

\begin{figure}[!thb] % Figure 1
\begin{center}
\includegraphics[width=7.0cm]{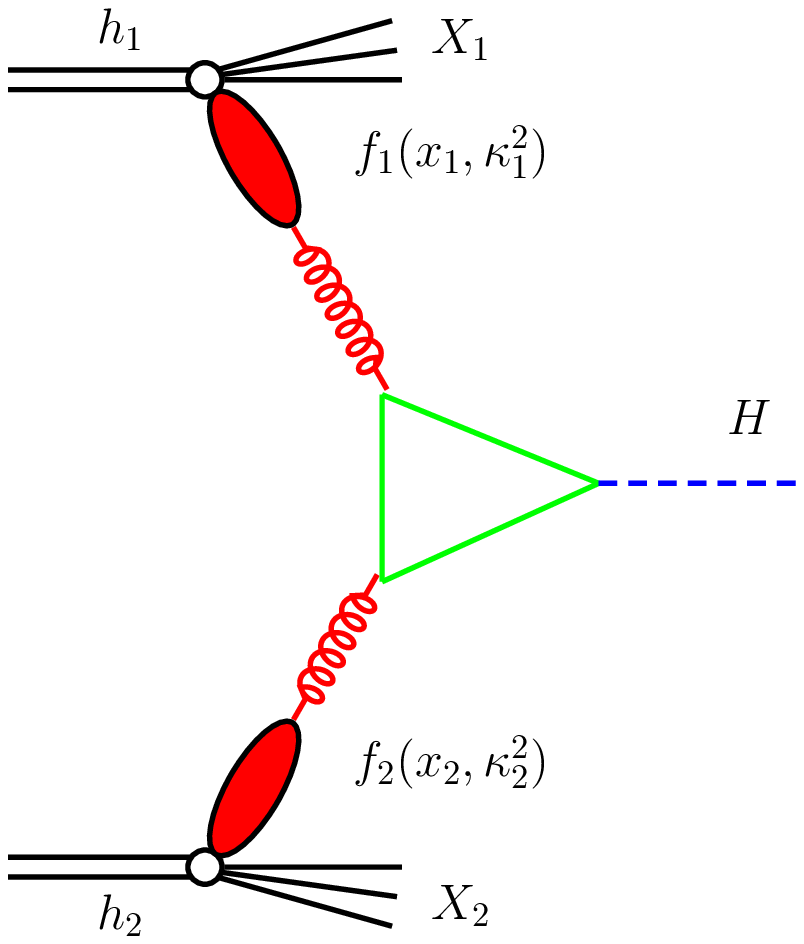}
\caption[*]{Dominant leading-order diagram for inclusive Higgs
production for $p_t \ll M_H$.
\label{fig:diagram}
}
\end{center}
\end{figure}

There are several mechanisms of the Higgs boson production
which have been discussed in the literature. 
Provided that the Higgs mass is larger than 100 GeV and smaller than
600 GeV the gluon-gluon fusion (see Fig.\ref{fig:diagram})
is the dominant mechanism of Higgs boson production at
LHC energies \cite{book}.
The WW and ZZ fusion is the second important mechanism for the
light-Higgs scenario.
Often the so-called associated Higgs boson production
($H W$, $H Z$ or $H t \bar t$) is considered as a good
candidate for the discovery of the Higgs boson. The contribution of
the associated production to the inclusive cross section is, however,
rather small. In the present paper we concentrate on the inclusive cross
section and in particular on its dependence on rapidity and/or
transverse momentum of the Higgs boson.

%-------------------------------------
\subsection{Gluon-gluon fusion}
%-------------------------------------

In the leading-order collinear factorization approach the Higgs boson
has a zero transverse momentum. In the collinear approach the finite
transverse momenta are generated only at next-to-leading order. However,
the fixed-order approach is not useful for small transverse momenta
and rather a resummation method must be used \cite{CSS}.
Recently Kwieci\'nski, starting from the CCFM equation \cite{CCFM},
has proposed a new method of resummation \cite{GK04} based
on unintegrated gluon distributions.

Limiting to small transverse momenta of the Higgs boson, i.e. small
transverse momenta of the fusing gluons, the on-shell approximation
for the matrix element seems a good approximation. There are also some
technical reasons, to be discussed later, to stay with the on-shell
approximation.
In the on-shell approximation for the $gg \to H$ transition matrix
element \footnote{There is a disagreement in the literature
\cite{hautmann02,LZ05} how to include
the off-shell effects.} the leading order cross section
in the unintegrated gluon distribution formalism reads
\footnote{Some of UGDF's from the literature depend only on longitudinal
momentum fraction and transverse momentum of the virtual gluon. To keep
formulae below general we allow for a scale parameter needed
in some distributions.} 
\begin{equation}
\frac{d \sigma^{H}}{dy d^2 p_t} = \sigma_0^{gg \to H}
\int
{  f_{g/1}(x_1,\kappa_1^2,\mu^2)} \;
{  f_{g/2}(x_2,\kappa_2^2,\mu^2)} \;
\delta^2(\vec{\kappa}_1+\vec{\kappa}_2 - \vec{p}_t)
\; \frac{d^2 \kappa_1}{\pi} \frac{d^2 \kappa_2}{\pi}    \; .
\label{momentum_representation_1}
\end{equation}
In the equation above the delta function assures conservation
of transverse momenta in the gluon-gluon fusion subprocess.
The 1/$\pi$ factors are due to the definition of UGDF's.
The momentum fractions should be calculated as
$x_{1,2} = \frac{m_{t,H}}{\sqrt{s}} \exp(\pm y)$, where in comparison
to the collinear case $M_H$ is replaced by the Higgs transverse mass
$m_{t,H}$.
If we neglect transverse momenta and perform the following formal
substitutions
\begin{equation}
\begin{split}
&f_{g/1}(x_1,\kappa_1^2,\mu^2) \to x_1 g_1(x_1,\mu^2) \;
\delta(\kappa_1^2) \; ,
\\
&f_{g/2}(x_2,\kappa_2^2,\mu^2) \to x_2 g_2(x_2,\mu^2) \;
\delta(\kappa_2^2) \; .
\end{split}
\label{kt_to_collinear}
\end{equation}
we recover the well known leading-order formula
\begin{equation}
\frac{d \sigma^{H}}{dy d^2 p_t} = \sigma_0^{gg \to H}
x_1 g_1(x_1,\mu^2) \; x_2 g_2(x_2,\mu^2) \; \delta^2(\vec{p}_t) \; .
\label{LO_collinear}
\end{equation}

The off-shell effects could be taken into account by inserting the
$\gamma^* \gamma^* \to H$ off-shell cross section (corresponding
to the matrix element squared
$|{\cal M}^{gg \to H}(\vec{\kappa}_1,\vec{\kappa}_2)|^2$)
under the integral in the formula
(\ref{momentum_representation_1}) above. This will be
discussed in more detail in a separate subsubsection.

There are a few conventions for UGDF in the literature.
In the convention used throughout the present paper the
unintegrated gluon distributions have dimension of GeV$^{-2}$
and fulfil the approximate relation
\begin{equation}
\int_{0}^{\mu^2} f_{g}(x,\kappa^2(,\mu^2)) \; d \kappa^2
 \approx x g_{coll}(x,\mu^2) \; ,
\end{equation}
where $g_{coll}(x,\mu^2)$ is the familiar conventional (integrated)
gluon distribution. The scale $\mu^2$ in UGDF above is optional.
In the effective Lagrangian approximation and assuming
infinitely heavy top quark the cross section parameter
$\sigma_0^{gg \to H}$ is given by \cite{effective_lagrangian}
\begin{equation}
\sigma_0^{gg \to H} = \frac{\sqrt{2} G_F}{576 \pi} \alpha_s^2(\mu_r^2)
\; .
\label{sigma_0}
\end{equation}
In the following we shall take $\mu_r^2 = M_H^2$.
Above we have assumed implicitly that the fusing gluons are
on-mass-shell. In general, the fusing gluons are off-mass-shell.
This effect was analyzed in detail in the production of Higgs
associated with two jets \cite{DKOSZ03}. The effect found there
is small provided $M_H < 2 m_t$.

The UGDF's are the main ingredients in evaluating the inclusive
cross section for Higgs production.
Depending on the approach, some UGDF's \cite{GBW_gluon,KL01,AKMS94,KuSt04}
in the literature depend on two variables
 -- longitudinal momentum fraction $x$ and transverse momentum $\kappa^2$,
some in addition depend on a scale parameter
\cite{KMR01,CCFM_b1,CCFM_b2,Bluemlein}. In the last case the scale
$\mu^2$ is taken here as $M_H^2$ or $\xi M_H^2$, where $\xi$
is some factor.

The seemingly 4-dimensional integrals in
Eq.(\ref{momentum_representation_1})
can be written as 2-dimensional integrals after a suitable change
of variables $\vec{\kappa}_1, \vec{\kappa}_2 \to \vec{p}_t, \vec{q_t}$,
where $\vec{p}_t = \vec{\kappa}_1 + \vec{\kappa}_2$ and
$\vec{q}_t = \vec{\kappa}_1 - \vec{\kappa}_2$. Then
\begin{equation}
\frac{d \sigma^{H}}{dy d^2 p_t} = \frac{\sigma_0^{gg \to H}}{(2 \pi)^2}
 \int \;
{  f_{g/1}(x_1,\kappa_1^2,\mu^2)} \;
{  f_{g/2}(x_2,\kappa_2^2,\mu^2)}
\; d^2 q_t   \; ,
\label{momentum_representation_2}
\end{equation}
where $\vec{\kappa}_1 = \vec{p}_t/2 + \vec{q}_t/2$ and
$\vec{\kappa}_2 = \vec{p}_t/2 - \vec{q}_t/2$.
The integrand of this ``reduced'' 2-dimensional integral in 
$\vec{q}_t = \vec{\kappa_1} - \vec{\kappa_2}$ is
generally a smooth function of $q_t$ and corresponding azimuthal
angle $\phi_{q_t}$.

%-----------------------------------------
\subsubsection{Unintegrated gluon distributions}
%-----------------------------------------

In the present analysis we shall use different unintegrated gluon distributions
from the literature. We include gluon distributions
corresponding to a simple saturation model used by Golec-Biernat and
W\"usthoff to describe the HERA deep-inelastic data \cite{GBW_gluon}
(GBW), a saturation model of Kharzeev and Levin
used to describe rapidity distributions of charged particles
\cite{KL01} (KL) \footnote{The normalization of the gluon distributions
was fixed in \cite{szczurek03} to reproduce the HERA data.},
Balitskij-Fadin-Kuraev-Lipatov (BFKL)-type UGDF \cite{AKMS94} and
three other distributions
in the transverse-momentum space \cite{Bluemlein},
\cite{KMR01} (KMR) and \cite{KuSt04} as well as the Kwieci\'nski
UGDF in the b-space. A more detailed description of almost
all distributions mentioned above can be found e.g. in
Ref.\cite{szczurek03}.
The Kwieci\'nski UGDF as the only one defined in the b-space requires
a separate discussion. It will be shown below that the formulae
for the inclusive cross section for Higgs boson production
via gluon-gluon fusion can be written in the equivalent way also in
terms of distributions in the b-space.

%------------------------------------------------------------------
\subsubsection{Kwieci\'nski gluon distribution and the inclusive cross section}
%------------------------------------------------------------------

Taking the following representation of the $\delta$ function
\begin{equation}
\delta^2(\vec{\kappa_1}+\vec{\kappa_2}-\vec{p}_t) =
\frac{1}{(2 \pi)^2} \int d^2 b \;
\exp \left[   
(\vec{\kappa_1}+\vec{\kappa_2}-\vec{p}_t) \vec{b}
\right] \; ,
\label{delta_representation}
\end{equation}
the formula (\ref{momentum_representation_2}) can be written in
the equivalent way in terms of gluon distributions in the space
conjugated to the gluon transverse momentum
\footnote{The simple form of the formula below would not be
possible with off-shell effects, i.e. when
$|{\cal M}^{gg \to H}|^2$ is a function of $\vec{\kappa}_1$
and $\vec{\kappa}_2$.}
\begin{equation}
\frac{d \sigma^{H}}{dy d^2 p_t} = \sigma_0^{gg \to H} \;
\int
{\tilde f}_{g/1}(x_1,b,\mu^2) \;
{\tilde f}_{g/2}(x_2,b,\mu^2)
 J_0(p_t b) \; 2 \pi b db    \; ,
\label{b_representation}
\end{equation}
where
\begin{equation}
{\tilde f}_g(x,b,\mu^2) = \int_0^{\infty} d \kappa_t \kappa_t
 J_0(\kappa_t b)
f_{g}(x,\kappa_t^2,\mu^2) \; .
\label{Fourier_transform}
\end{equation}
For most of unintegrated gluon distributions formula
(\ref{momentum_representation_2}) is used.
Formula (\ref{b_representation}) is used when applying
the Kwieci\'nski unintegrated distributions obtained as a solution
of his equations in the b-space.
The b-space approach proposed by Kwieci\'nski is very convenient to
introduce the nonperturbative effects like
intrinsic (nonperturbative) transverse momentum distributions
of partons in nucleons.
It seems reasonable, at least in the first approximation,
to include the nonperturbative effects in the factorizable way
\begin{equation}
\tilde{f}_g(x,b,\mu^2) = 
\tilde{f}_g^{CCFM}(x,b,\mu^2)
 \cdot F_g^{np}(b) \; .
\label{modified_uPDFs}
\end{equation}
The form factor responsible for the nonperturbative effects
must be normalized such that
\begin{equation}
F_g^{NP}(b=0) = 1 \; .
\label{ff_normalization}
\end{equation}
Then by construction:
\begin{equation}
{\tilde f}_g(x,b=0,\mu^2) = \frac{x}{2} g(x,\mu^2) \; .
\end{equation}
In the following, for simplicity, we use an
$x$-independent form factor
\begin{equation}
F_g^{np}(b) = \exp\left(-\frac{b^2}{4 b_0^2}\right) \; 
\label{formfactor}
\end{equation}
which is responsible for the nonperturbative effects.
The Gaussian form factor in $b$ means also a Gaussian initial
momentum distribution $\exp(-k_t^2 b_0^2)$ (Fourier transform of
a Gaussian function is a Gaussian function). Gaussian form factor
is often used to correct collinear pQCD calculations for so-called
intrinsic momenta. Other functional forms in $b$ are also possible.

The similarities and differences betwen the standard soft gluon
resummation and the CCFM resummation have been discussed recently
in Ref.\cite{GK04}.
It has been shown how the soft gluon resummation formulae can be
obtained as the result of the approximate treatment of the solution of
the CCFM equation in the so-called b-representation.

%------------------------------------------------------
\subsubsection{Off-shell matrix element for $g^* g^* \to H$}
%------------------------------------------------------

While in the collinear approach the off-shell matrix elements
are needed only at higher orders \cite{hautmann02},
in the $k_t$-factorization approach the off-shell matrix elements
appear in principle already at the leading order.
In Ref.\cite{hautmann02} the matrix element was calculated
in the framework of the effective Lagrangian for the Higgs
boson coupling to gluons in the infinitely heavy top mass approximation.
While the on-mass-shell couplings in the full theory and
in the effective Lagrangian theory are equivalent, this is not
expected for matrix element with off-shell gluons. In particular
the dependence on transverse momenta and their relative orientation
can be different.
The use of the effective vertex all over the phase space may not
be completely realistic.
However, the matrix element in the full theory (with finite top mass)
was not yet calculated.

In the present paper we shall only estimate the
off-shell effects in the effective Lagrangian approximation.
Then the cross section for the Higgs production can be written as:
\begin{equation}
\frac{d \sigma^{H}}{dy d^2 p_t} = \frac{\sigma_0^{gg \to H}}{(2 \pi)^2}
 \int \;
{  f_{g/1}(x_1,\kappa_1^2,\mu^2)} \;
{  f_{g/2}(x_2,\kappa_2^2,\mu^2)} \;
2 \frac{({\vec \kappa}_{1} \cdot {\vec \kappa}_{2})^2}
       {\kappa_{1}^2 \kappa_{2}^2} 
\lambda(\kappa_{1}^2,\kappa_{2}^2,p_t^2)
\; d^2 q_t   \; ,
\label{Higgs_off_shell}
\end{equation}
where $\lambda$ is a smooth function of its parameters
\cite{hautmann02,LZ05}.
In the limit $p_t^2 \to 0$, $\kappa_1^2 \to 0$,
$\kappa_2^2 \to 0$ the dimensionless function
$\lambda(\kappa_{1}^2,\kappa_{2}^2,p_t^2) \to 1$.
Then the 2 $\cos^2\phi_{\vec{\kappa}_{1},\vec{\kappa}_{2}}$
factor in the formula above constitute the essential difference
with respect to the on-shell approximation. It modifies the
dependence of the integrand under the $\phi_q$ integration.

\begin{figure}[!thb] % Figure 1a
\begin{center}
\includegraphics[width=10.0cm]{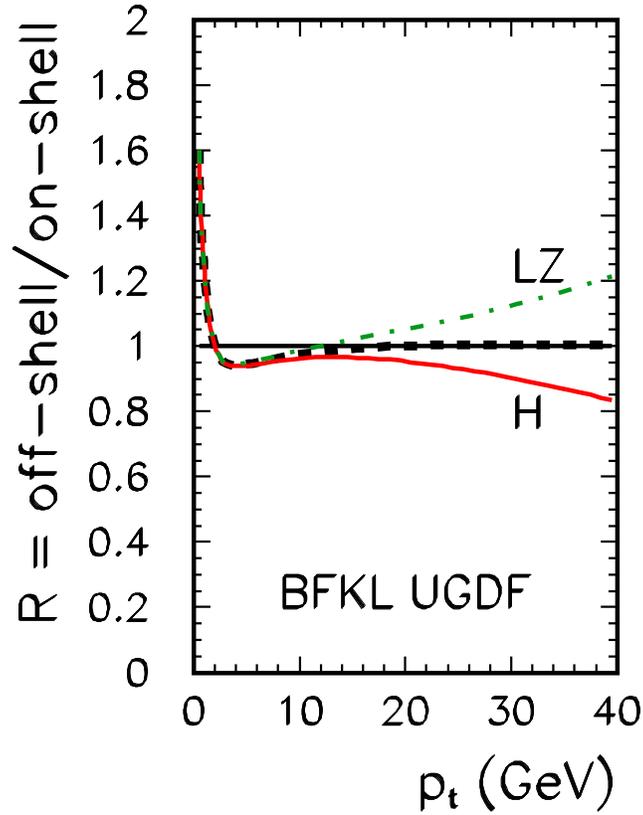}
\caption[*]{
The ratio $R$ as a function of Higgs transverse momentum for
the full range of Higgs rapidity.
The thick dashed line corresponds to neglecting the function
$\lambda$, i.e. assuming $\lambda$ = 1. The solid line was
obtained with the Hautmann prescription of the flux factor
\cite{hautmann02} while the dash-dotted line is based
on the formula from Ref.\cite{LZ05}. In this calculation the BFKL
UGDF was used as an example.
\label{fig:offshelltoonshell}
}
\end{center}
\end{figure}

In Fig.\ref{fig:offshelltoonshell} we show the ratio:
\begin{equation}
R = \frac{(d\sigma/dp_t)_{off-shell}}{(d\sigma/dp_t)_{on-shell}}
\label{off-shell-on-shell-ratio}
\end{equation}
as a function of Higgs-boson transverse momentum.
In this calculation the BFKL unintegrated gluon distributions
have been used as an example. The result for other distributions
is similar. The final effect is rather small (less than 10\%)
in the region of our interest and depends on the way how
the flux factor for off-shell gluons is defined \cite{hautmann02,LZ05}
\footnote{While it is rather strightforward to calculate matrix
  element for off-shell gluons, there is some ambiguity in defining
 the flux factor for virtual gluons.}.
At finite, but not too large, Higgs boson transverse momenta
the averaging over gluon transverse momenta with UGDF's gives
$< 2 cos^2\phi_{\kappa_{1},\kappa_{2}} > \approx$ 1, and one
approximately recovers the on-shell result. This is not true
for $p_t \approx$ 0, when $\vec{\kappa_1}$ and $\vec{\kappa_2}$
are strongly anticorrelated and the averaging is not efficient.
The off-shell effect on the integrated cross section ($d\sigma /dy$)
is even smaller.
Here we wish to concentrate rather on the effect of transverse momenta
inherent for UGDF's.
We shall leave a detailed study of the off-shell effects
in effective Lagrangian and in the full theory for
the future and consequently we shall use the on-shell matrix element
in the following.
This approximation will be also useful here when comparing
the $k_t$-factorization results with that for the standard collinear
and soft-gluon resummation approaches.
The on-shell approximation was used recently in the formalism
od doubly unintegrated parton distribution for electroweak boson
production \cite{WMR04}.

%-------------------------------------------
\subsubsection{Standard soft-gluon resummation}
%-------------------------------------------

The formula for inclusive cross section in terms of unintegrated
gluon distributions in the impact parameter space looks very similar
to the one in the standard soft gluon resummation approach known
from the literature. This similarity is not random \cite{GK04}.
In the Collins-Soper-Sterman (CSS) approach \cite{CSS} the resummed
cross section for Higgs production reads
\begin{equation}
\begin{split}
\frac{d \sigma}{d y d^2 p_{t,H}} =
\frac{\sigma_0^{gg \to H}}{(2 \pi)^2} 
\int d^2 b \; J_0(p_t b) \; W^{NP}_{g g}(b,x_1,x_2,\mu^2) \\
 x_1 \cdot \left[ g_1(x_1,\mu(b)) 
       + \frac{\alpha_s(\mu(b))}{2 \pi} C_{vc} g_1(x_1,\mu(b))
       + \frac{\alpha_s(\mu(b))}{2 \pi}
\sum_{f_1}(C_{g q} \otimes q_{1}^{f_1})(x_1,\mu(b))
   \right]  \\
 x_2 \cdot \left[ g_2(x_2,\mu(b)) 
       + \frac{\alpha_s(\mu(b))}{2 \pi} C_{vc} g_2(x_2,\mu(b))
       + \frac{\alpha_s(\mu(b))}{2 \pi}
\sum_{f_2}(C_{g q} \otimes q_{2}^{f_2})(x_2,\mu(b))
   \right]  \\
\exp\left[ \frac{1}{2} \left(S_g(b,\mu^2) + S_g(b,\mu^2) \right)
\right] \; ,
\label{CSS_formula}
\end{split}
\end{equation}
where the exponents in the Sudakov-like form factors read
\begin{equation}
S_g(b,\mu^2) =
- \int_{\bar \mu_{min}^2(b)}^{\mu^2} \frac{d \bar \mu^2}{\bar \mu^2}
\left[
\ln \left(\frac{\mu^2}{\bar \mu^2} \right) A_g(\alpha_s(\bar \mu^2)) +
                                        B_g(\alpha_s(\bar \mu^2))
\right] \; .
\label{Sudakov}
\end{equation}
The coefficient functions $C$'s in Eq.(\ref{CSS_formula}) can be
found in \cite{resummation}.
The coefficient $A$ and $B$ in the Sudakov-like form factor
can be expanded in the series of $\alpha_s$:
\begin{eqnarray}
A_g &=& 2 C_A \frac{\alpha_s(\bar \mu)}{2 \pi}
     +  \left(\frac{\alpha_s(\bar \mu)}{2 \pi}\right)^2 (...) + ... \; ,
 \nonumber
\\
B_g &=& -2 \beta_0 \frac{\alpha_s(\bar \mu)}{2 \pi}
     +  \left(\frac{\alpha_s(\bar \mu)}{2 \pi}\right)^2 (...) + ... \; ,
\label{expansion}
\end{eqnarray}
where $\beta_0 = \frac{11}{6} C_A - \frac{2}{3} N_F T_R$
 ($T_R = \frac{1}{2}, N_F = 5, C_A = 3$).
The CSS formalism \cite{CSS} leaves open the question of small $b$.
Different prescriptions have been proposed to treat this region.
The lower limit of the integral in Eq.(\ref{Sudakov}) is usually taken
$\mu^2_{min}(b) = \left( \frac{C_b}{b} \right)^2$, where $C_b$ = 2
$\exp( -\gamma_E ) \approx$ 1.1229. This prescription leads
to a kink for the Sudakov form factor if $C_b/b = \mu$.
To allow a smooth dependence and to quarantee that the lower limit
is really lower than the upper limit, one could make the following
replacement
$\mu^2_{min}(b) = \left( \frac{C_b}{b} \right)^2 \rightarrow 
\left( \frac{C_b}{b} \right)^2
\left[1 + C_b^2/(b^2 \mu^2) \right]^{-1}$.
To quarantee that the scale of parton distribution does not take
unphysically small value we shall use the following prescription:
\begin{equation}
\mu^2(b) = \mu^2_{min}(b) + \mu_0^2 \; ,
\label{PDF_scale}
\end{equation}
where $\mu_0^2$ is the starting value for the QCD evolution.
In the present paper we shall use easy to handle leading
order parton distributions from Ref.\cite{GRV98}.

$W_{gg}^{NP}(b,x_1,x_2,\mu^2)$ in Eq.(\ref{CSS_formula}) is
of nonperturbative origin.
Different effective parametrizations have been proposed in
the literature.
Assuming a factorizable form of the $W_{gg}^{NP}$ function
\begin{equation}
W_{gg}^{NP}(b,x_1,x_2,\mu^2) = F_g^{NP}(b,x_1,\mu^2) \cdot
 F_g^{NP}(b,x_2,\mu^2)
\end{equation}
the soft-gluon-resummation formula (\ref{CSS_formula}) and
the unintegrated gluon distribution formula (\ref{b_representation})
for Higgs production in the b space have identical structure if
the following formal assignment is made:
\begin{equation}
{\tilde f}_g^{SGR}(x,b,\mu^2) = \frac{1}{2} F_g^{NP}(b,x,\mu^2)
[ \; x g(x,\mu^2(b)) + ...] \exp \left( \frac{1}{2} S_g(b,\mu^2) \right)
\; .
\label{formal_SGR_UPDF}
\end{equation}

If the off-shell matrix element for $gg \to H$ is taken in the UGDF approach
with the Kwieci\'nski distribution, the structure of the formula
in both approaches would be different. In this sense the UGDF approach
seems more general than the b-space resummation method.

%-------------------------------------
\subsection{2 $\to$ 2 processes}
%-------------------------------------

At sufficiently large transverse momenta ($p_t > M_H$) the Higgs boson
production of the type $2 \to 2$ should dominate over the $2 \to 1$
mechanism discussed above.  
The cross section for fixed-order processes of the type
$p_1 p_2 \to H p_3$ (parton+parton $\to$ Higgs+parton)
of the order of $\alpha_s$ is well known \cite{2to2_basis}
\begin{equation}
  \begin{split}
\frac{d \sigma}{dy_H dy_p d^2 p_t}(y_W,y_p,p_t) &=
\frac{1}{16 \pi^2 {\hat s}^2 }  \\
&\times \biggl\{ x_1 g_1(x_1,\mu^2) \; x_2 g_2(x_2,\mu^2) \;
 \overline{|{\cal M}_{gg \to Hg}|^2}   \\ 
&+ \left[ \sum_{{f_1}=-3,3} x_1 q_{1,f_1}(x_1,\mu^2) \right] \; 
x_2 g_2(x_2,\mu^2) \;
 \overline{|{\cal M}_{qg \to Hq}|^2} \;  \\
&+x_1 g_1(x_1,\mu^2) \; 
 \left[ \sum_{{f_2}=-3,3} x_2 q_{2,f_2}(x_2,\mu^2) \right]       
 \overline{|{\cal M}_{gq \to Hq}|^2} \;  \\
&+ \sum_{f=-3,3} x_1 q_{1,f}(x_1,\mu^2) \; x_2 q_{2,-f}(x_2,\mu^2)
 \overline{|{\cal M}_{qq \to Hg}|^2} \; \biggr\} \; .   
\label{2to2}
  \end{split}
\end{equation}
The indices $f$ in the formula above number both quarks ($f >$ 0)
and antiquarks ($f <$ 0). Only three light flavours are included in
actual calculations.
The explicit formulae for $\overline{|{\cal M}|^{2}}$ can be found
in \cite{2to2_basis}.

%-------------------------------------
\subsection{Weak boson fusion}
%-------------------------------------

Up to now we have discussed only the contribution of the dominant
LO gluon-gluon fusion and NLO 2 $\to$ 2 corrections and completely
ignored contributions of other processes.
The second most important mechanism for Higgs production
is the fusion of off-shell gauge bosons: WW or ZZ. It is known
that at LHC energy and intermediate mass (100 GeV $< M_H <$ 500 GeV)
Higgs the WW fusion constitutes about 10-15 \% of the integrated
inclusive cross section. If the weak boson fusion contribution
was separated, the measurement of the WWH (or ZZH) coupling
would be very interesting test of the Standard Model.

Previous studies of the WW mechanism concentrated on the
total cross section for the Higgs production. 
In the present paper we are interested in differential distributions
of Higgs boson rather than in the integrated cross section. 

For the gauge boson fusion the partonic
subprocess is of the 2 $\to$ 3 type:
$q(p_1) + q(p_2) \to q(p_3) + q(p_4) + H(p_H)$. 
The corresponding hadronic cross section can be written as
\begin{equation}
\begin{split}
d \sigma = {\cal F}_{12}^{VV}(x_1,x_2) \; \frac{1}{2 \hat s}
\; \overline{ | {\cal M}_{qq \to qqH} |^2 } \;
\frac{d^3 p_3}{(2 \pi)^3 2 E_3}
\frac{d^3 p_4}{(2 \pi)^3 2 E_4}
\frac{d^3 p_H}{(2 \pi)^3 2 E_H} \\
(2 \pi)^4 \delta^{4}(p_1+p_2-p_3-p_4-p_H) \; d x_1 d x_2 \; .
\end{split}
\label{WW_fusion}
\end{equation}
The next-to-leading order corrections to the matrix element of
the WW fusion are rather small \cite{WW_NLO}.
For comparison the NLO corrections for gluon-gluon fusion are
significantly larger. Since we wish to concentrate on relative
effect of the gluon-gluon and WW fusion
contributions in the following we restrict to a much simpler
leading order (LO) calculation.
The LO subprocess matrix element was calculated first in
Ref.\cite{CD84}.
The spin averaged matrix element squared reads
\begin{equation}
\overline{ | {\cal M} |^2 } = 128 \sqrt{2} G_F^3
\frac{M_W^8 (p_1 \cdot p_2) (p_3 \cdot p_4)}
{(2 p_3 \cdot p_1 + M_W^2)^2 (2 p_4 \cdot p_2 + M_W^2)^2} \; .
\label{M_WW}
\end{equation}
For the WW fusion, limiting to light flavours,
the partonic function is
\begin{equation}
\begin{split}
{\cal F}_{12}^{WW}(x_1,x_2) = \\
\left( u_1(x_1,\mu_1^2)+\bar d_1(x_1,\mu_1^2)+\bar s_1(x_1,\mu_1^2) \right)
\left( \bar u_2(x_2,\mu_2^2)+d_2(x_2,\mu_2^2)+s_2(x_2,\mu_2^2) \right) + \\
\left( \bar u_1(x_1,\mu_1^2)+d_1(x_1,\mu_1^2)+s_1(x_1,\mu_1^2) \right)
\left( u_2(x_2,\mu_2^2)+\bar d_2(x_2,\mu_2^2)+\bar s_2(x_2,\mu_2^2) \right)
\; .
\end{split}
\label{partonic_function}
\end{equation}
We take either (i)  $\mu_1^2 = \mu_2^2 = M_H^2$ or 
(ii) $\mu_1^2 = -t_1, \mu_2^2
= -t_2$, where $t_1$ and $t_2$ are virtualities of W bosons.
It is convenient to introduce the following new variables:
\begin{equation}
\begin{split}
\vec{p}_{+} = \vec{p}_{3} + \vec{p}_{4} \; , \\
\vec{p}_{-} = \vec{p}_{3} - \vec{p}_{4} \; ,
\end{split}
\label{pplus_pminus}
\end{equation}
which allow to eliminate the momentum-dependent $\delta^3(...)$
in Eq.(\ref{WW_fusion}).
Instead of integrating over $x_1$ and $x_2$ we shall
integrate over
$y_1 \equiv \ln(1/x_1)$ and $y_2 \equiv \ln(1/x_2)$.
Then using Eq.(\ref{WW_fusion}) we can write the inclusive
spectrum of Higgs as
\begin{equation}
\begin{split}
\frac{d \sigma}{dy d^2 p_t} =& \int d y_1 d y_2 \;
x_1 x_2 {\cal F}(x_1,x_2,\mu_1^2,\mu_2^2)
\; \frac{1}{2 \hat s} \frac{d^3 p_{-}}{16}
\; \overline{ | {\cal M}_{qq \to qqH} |^2 } \;
\frac{1}{2E_3} \; \frac{1}{2E_4} \; \\
&\frac{1}{(2 \pi)^5} \; \delta(E_1+E_2-E_3-E_4-E_H) \; .
\end{split}
\label{WW_fusion_red1}
\end{equation}
This is effectively a four-dimensional integral which
can be easily calculated numerically.

%----------------
\section{Results}
%----------------

%-------------------------------------
\subsection{Gluon-gluon fusion}
%-------------------------------------

Since we wish to concentrate on the potential to verify different
UGDF's rather than to present the best predictions for LHC experiments
we shall consider only one mass of the Higgs boson $M_H$ = 125 GeV
as an example.
This is slightly above the lower limit obtained from the analysis
of the LEP data \cite{LEP_limit}.
In addition, this is a mass for which many calculations in
the literature has been performed recently. Therefore this gives
a chance of a comparison to the existing results.

\begin{figure}[htb] % Figure 2
\begin{center}
    \includegraphics[width=6.5cm]{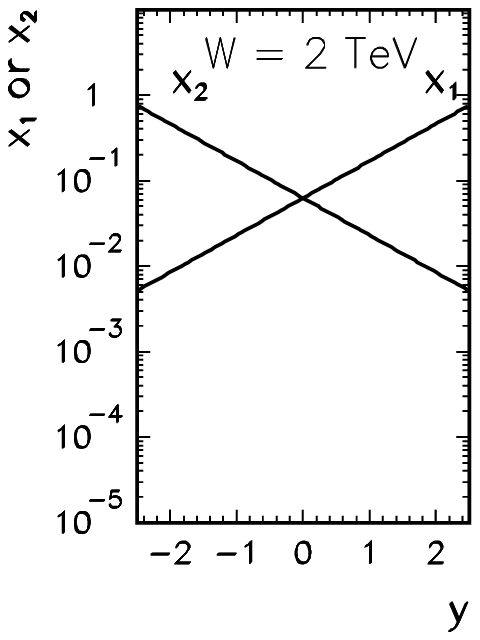}
    \includegraphics[width=6.5cm]{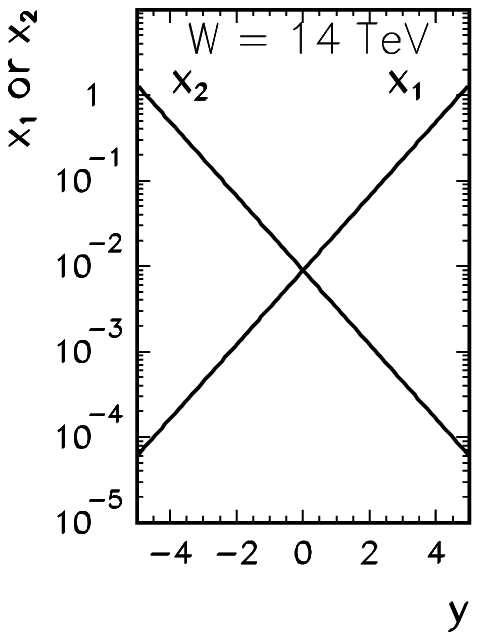}
\caption{\it
$x_1$ and $x_2$ as a function of Higgs rapidity $y$ for $p_t$ = 0. 
In panel (a) for Tevatron energy $\sqrt{s}$ = 2 TeV and
in panel (b) for LHC energy $\sqrt{s}$ = 14 TeV.
\label{fig:x1x2_y}
}
\end{center}
\end{figure}

Before we go to the analysis of the two-dimensional spectra of
Higgs boson produced in proton-proton or proton-antiproton collisions
let us show the range of gluon longitudinal momentum fraction
tested in these proceses. In Fig.\ref{fig:x1x2_y} we present
correspondingly $x_1$ and $x_2$ as a function of Higgs boson rapidity
for a few different values of Higgs transverse momentum.
While at Tevatron energies (panel (a)) only intermediate and large
$x$'s come into the game, in collisions at LHC energy (panel (b))
x$\sim$ 10$^{-2}$ is sampled at midrapidity. However, at LHC energy,
at rapidities $|y| >$ 2 one enters the region of $x >$ 0.1. Here some
of the low-x models of UGDF's may become invalid.

\begin{figure}[htb] % Figure 3
\begin{center}
    \includegraphics[width=10cm]{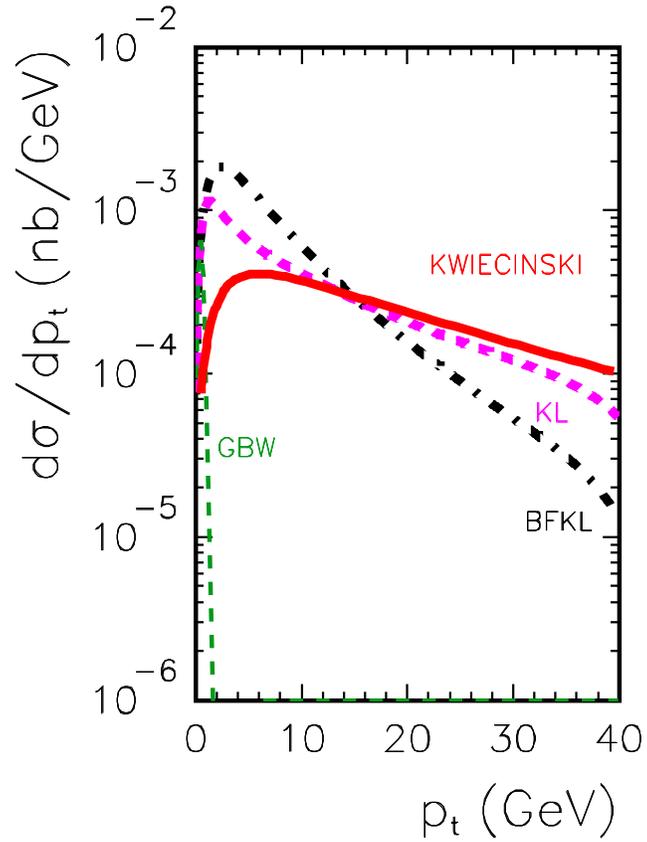}
\caption{\it
Transverse momentum distribution of Higgs boson at LHC energy
W = 14 TeV for different UGDF's from the literature:
solid -- Kwieci\'nski, thick dashed -- KL, thin dashed -- GBW,
dash-dotted -- BFKL. 
\label{fig:LHC_pt}
}
\end{center}
\end{figure}

\begin{figure}[htb] % Figure 4
\begin{center}
    \includegraphics[width=6.5cm]{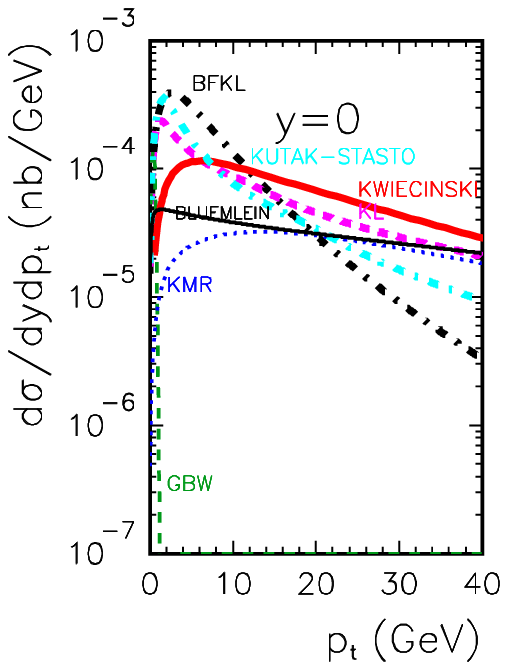}
    \includegraphics[width=6.5cm]{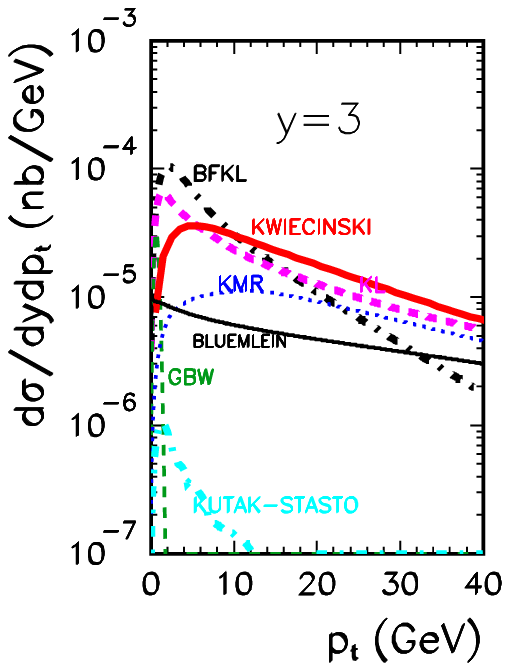}
\caption{\it
Transverse momentum distribution of Higgs boson at LHC energy
W = 14 TeV and y = 0 (left panel) and y = $\pm$ 3 (right panel)
for different UGDF's from the literature. The notation here is the
same as in the previous figure. In addition to the previous figure
we present results for KMR (dotted), Kutak-Sta\'sto (grey
dash-dotted) and Bluemlein (thin solid) UGDF.
\label{fig:LHC_pt_bins}
}
\end{center}
\end{figure}

Let us concentrate first on transverse momentum distributions.
The distribution of Higgs transverse momentum (rapidity integrated)
is shown in Fig.\ref{fig:LHC_pt}.
In Fig.\ref{fig:LHC_pt_bins} we present transverse momentum
distribution of Higgs boson in different bins of rapidity specified
in the figure caption. At midrapidity (panel (a)) 
only small $x$'s are sampled. Even here different models from the
literature give quite different transverse momentum distributions
although all of them give a reasonable description of the HERA data.
The LO soft-gluon resummation distribution and the distribution
obtained with the Kwieci\'nski unintegrated gluon distribution
($b_0$ = 1 GeV$^{-1}$)
are very similar with maxima at $p_{t,H} \approx$ 10 GeV
and $p_{t,H} \approx$ 5 GeV, respectively.
The cross section for W or Z production is fairly sensitive to
the choice of the nonperturbative form factor \cite{KS04}.
In contrast to the gauge boson production the Higgs boson
production is much less sensitive to the parameter of the Gaussian
form factor. The results with $b_0$ = 0.5 or 2 GeV$^{-1}$ (not shown here)
almost coincide with the result for $b_0$ = 1 GeV$^{-1}$.
Therefore in the following in all calculations we shall use
$b_0$ = 1 GeV$^{-1}$.
The BFKL type gluon distributions lead to much larger cross section
at small Higgs transverse momenta and a sizeably larger slope
of the $p_t$-distribution.
The rapid fall-off of the cross section for the Golec-Biernat-W\"usthoff
nonperturbative gluon distribution \cite{GBW_gluon} (thin dashed curve)
demonstrates how important is the perturbative initial state radiation
in generating larger transverse momenta of the Higgs boson.
Such effects are not taken into account in Ref.\cite{GBW_gluon}.

A comment regarding the GBW distribution is here in order.
This distribution was obtained based on a simple dipole
parametrization of the HERA data inspired by the saturation idea.
In addition to the very steep distribution in Higgs-boson transverse
momentum the scale-independent GBW gluon distribution gives very small
total cross section (about half of 1 pb).
One has to remember, however, that this distribution was constructed in
order to describe $\sigma_{\gamma^*p}^{tot}(Q^2)$ for small photon
virtualities.
We wish to stress here that this distribution is not an universal object.
For example in its simplest form it fails to describe
$\sigma_{\gamma^*p}^{tot}(Q^2)$ for large photon virtualities. 
The corresponding effective gluon distribution defined as
\begin{equation}
x g_{GBW}(x) \equiv \int_0^{\infty} \; {\cal F}_g^{GBW}(x) d \kappa^2
\label{effective_glue}
\end{equation}
resembles the standard collinear distribution
$x g_{DGLAP}(x,\mu^2)$ for small factorization scale $\mu^2 \sim$ 1
GeV$^2$. The latter, when substituted into the standard leading-order
formula, also leads to a small total cross section of the order of 1 pb.
A reasonable total cross section is obtained provided
$\mu^2 \sim M_H^2$. Although $g_{DGLAP}(x,\mu^2=1 GeV^2)$
leads to a reasonable description of $F_2$ at $Q^2 \sim$ 1 GeV$^2$
it cannot be directly (without DGLAP evolution) used for Higgs production.
In this context the scale-independent GBW distribution should be
understood as an initial condition for QCD evolution rather than
an universal object to be used in different high-energy processes.

Different UGDF's constructed in order to describe the total cross
section for $\gamma^* p$ process give quite different predictions
for the Higgs production.
This, as will be discussed below, is not completely surprising.
While the $\gamma^* p$ total cross section is sensitive to the integral
$\int d k_t^2 f_g(x,k_t^2(,\mu^2))$, the Higgs boson transverse momentum
distribution samples details of the unintegrated gluon distributions.
As discussed in the previous section, in the case of Higgs boson
production the inclusive cross section is a convolution of two UGDF's.
In general, exclusive reactions are much better place for testing UGDF's.
In this sense the standard procedure to constrain UGDF's through
describing the $F_2$ HERA data does not seem very effective.

\begin{figure}[htb] % Figure 5
\begin{center}
    \includegraphics[width=6.0cm]{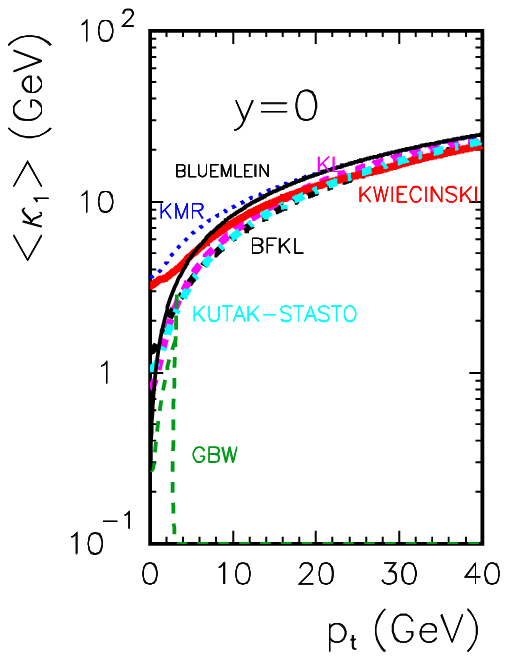}
    \includegraphics[width=6.0cm]{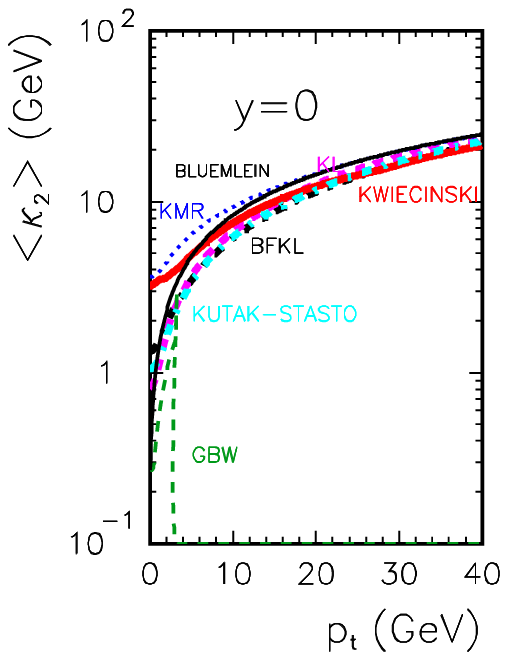}
    \includegraphics[width=6.0cm]{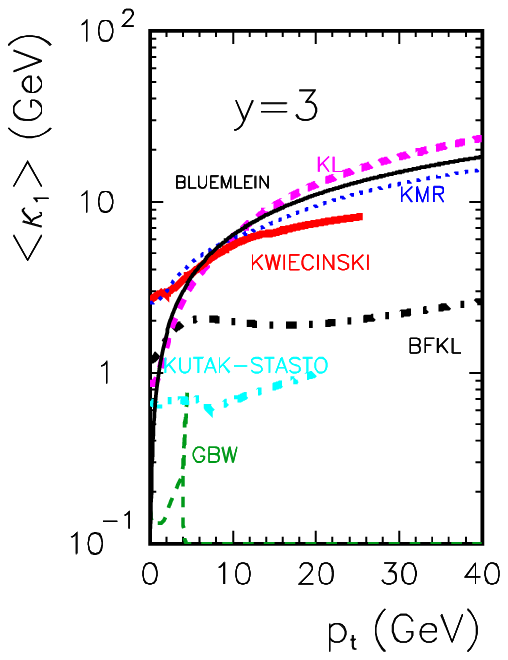}
    \includegraphics[width=6.0cm]{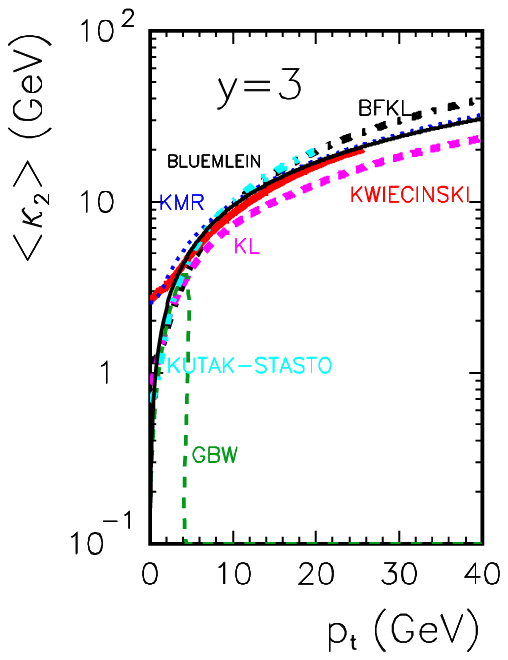}
\caption{\it
Average gluon transverse momentum as a function of Higgs transverse
momentum at LHC energy W = 14 TeV 
for different UGDF's from the literature.
\label{fig:LHC_kappa_pt_bins}
}
\end{center}
\end{figure}

In order to understand the situation somewhat better in
Fig.\ref{fig:LHC_kappa_pt_bins} we show the corresponding average
values of sampled transverse momenta as the function of Higgs transverse
momentum. At y=0, by symmetry requirement, $< \kappa_1 >$
and $< \kappa_2 >$ are identical. It is not the case for y=3.
Rather similar results are obtained with different UGDF's.
At very small Higgs transverse momenta one tests $\kappa$'s $\sim$ 1
GeV. At large Higgs transverse momenta and y=0 we get
$< \kappa_1 > + < \kappa_2 > \approx p_t/2$.
Of course by symmetry $< \kappa_{1/2} >(-y) = < \kappa_{2/1} >(y)$.

\begin{figure}[htb] % Figure 6
\begin{center}
    \includegraphics[width=10.0cm]{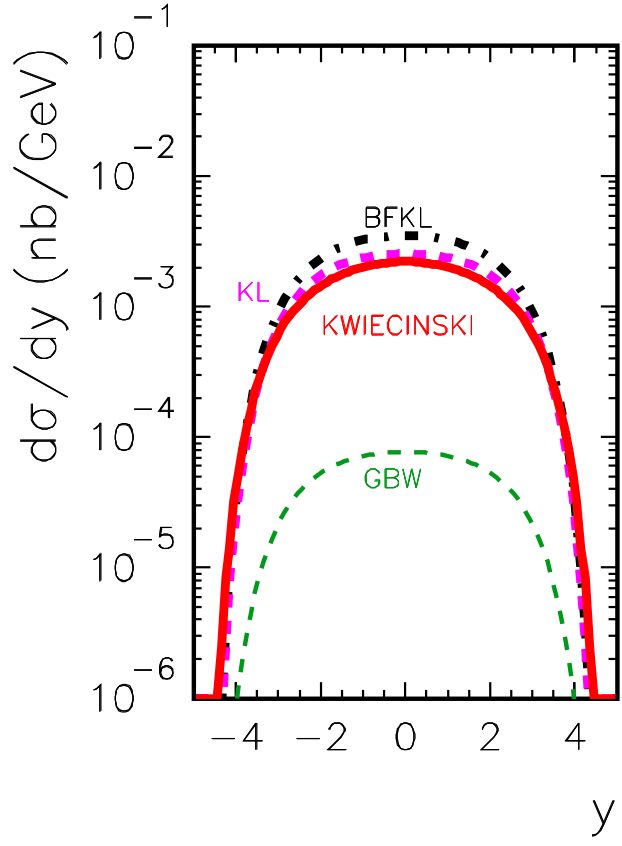}
\caption{\it
Rapidity distribution of Higgs at LHC energy W = 14 TeV for
different UGDF's from the literature.
\label{fig:LHC_y}
}
\end{center}
\end{figure}

Some examples of inclusive rapidity distributions are shown
in Fig.\ref{fig:LHC_y}.
Even here the differences between different UGDF's are clearly visible.
Above $|y| >$ 3 only resummation distribution (thin solid line),
Kwieci\'nski distribution (thick solid line) and
Kimber-Martin-Ryskin (dotted line) are applicable by construction.
The other distributions were obtained by extending the generally
small-x gluon distributions above x $>$ 0.1 by multiplying the
original formulae by $(1-x)^n$. The power n=5-7 was found recently
in the production of $c \bar c$ pairs in photon-proton
scattering at low energies \cite{LS04}. In the present paper we used
n=7. Small differences may be expected only at the very edges of the phase
space, i.e. in the region we are not interested here.

\begin{figure}[htb] % Figure 7
\begin{center}
    \includegraphics[width=6.5cm]{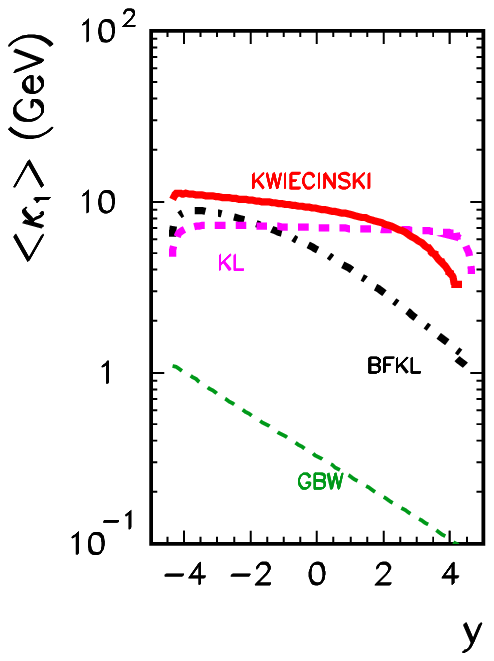}
    \includegraphics[width=6.5cm]{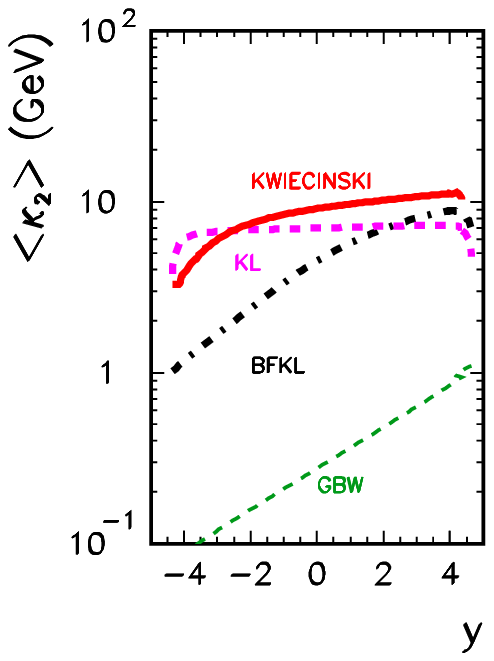}
\caption{\it
Average gluon transverse momentum $< \kappa_1 >$ or $< \kappa_2 >$
as a function of Higgs rapidity
for different UGDF's from the literature. In this calculation
$p_t <$ 40 GeV.
\label{fig:LHC_kappa_y}
}
\end{center}
\end{figure}

For completeness in Fig.\ref{fig:LHC_kappa_y} we show the average values
of gluon transverse momenta $<\kappa_1>$ and $<\kappa_2>$
as a function of Higgs-boson rapidity.
At midrapidity by symmetry $<\kappa_1> \approx <\kappa_2>$.
The average values strongly depend on UGDF used and on the region
of rapidity. The asymmetry of average values of the transverse momenta
at forward/backward regions are closely related to Fig.\ref{fig:x1x2_y}
due to correlation of transverse momenta with longitudinal momentum
fractions. Generally, the smaller $x_1$ ($x_2$) the larger $<\kappa_1>$
($<\kappa_2>$). The details depend, however, on the specific version of
the gluon dynamics.  
The extremely small average transverse momenta for
the GBW UGDF can be understood in the light of the discussion above.

\begin{figure}[htb] % Figure 8
\begin{center}
    \includegraphics[width=6.5cm]{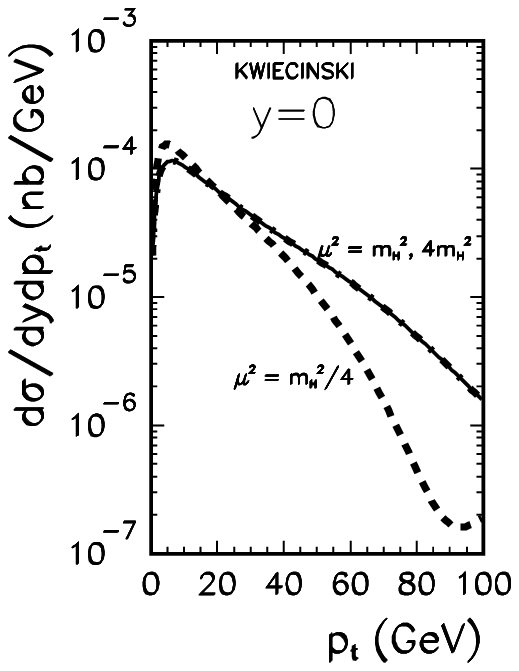}
    \includegraphics[width=6.5cm]{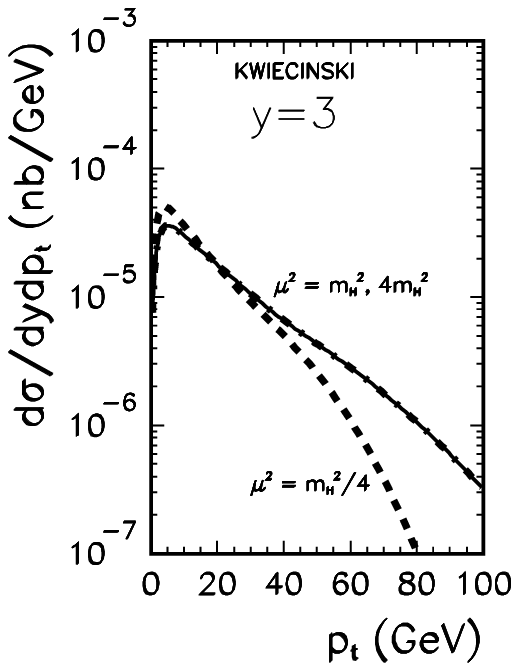}
\caption{\it
A dependence of the result on the choice of the factorization
scale for the Kwieci\'nski UGDF for y = 0 (left panel)
and y = 3 (right panel). In this calculation a Gaussian form factor
with $b_0 = 1 GeV^{-1}$ was used.
\label{fig:kwiec_factorization_scale}
}
\end{center}
\end{figure}

How important is the choice of the factorization scale in our
calculations with Kwieci\'nski UGDF?
In Fig.\ref{fig:kwiec_factorization_scale}
we show results obtained with quite different choices of factorization
scale and with $b_0$ = 1GeV$^{-1}$.
There is very little effect if the factorization scale
is increased from our cannonical value $\mu^2 = M_H^2$.
There is, however, a sizeable effect if the factorization scale
is decreased drastically, especially at $p_t >$ 50 GeV.
The factorization scale dependence in our case seems somewhat weaker
than in a recent work \cite{LZ05}.

\begin{figure}[htb] % Figure 9
\begin{center}
    \includegraphics[width=9cm]{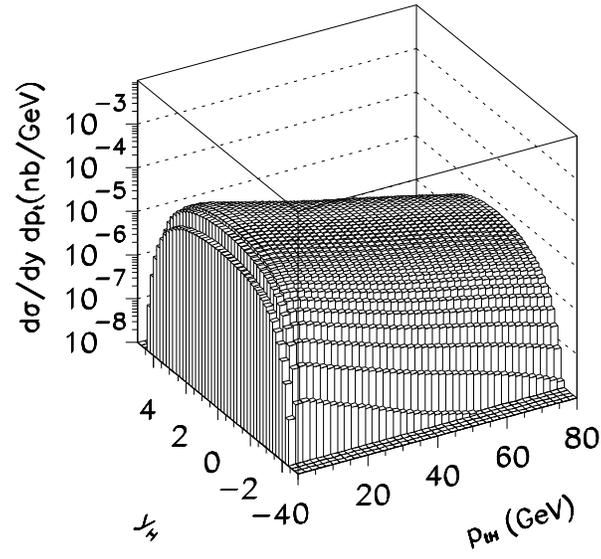}
    \includegraphics[width=9cm]{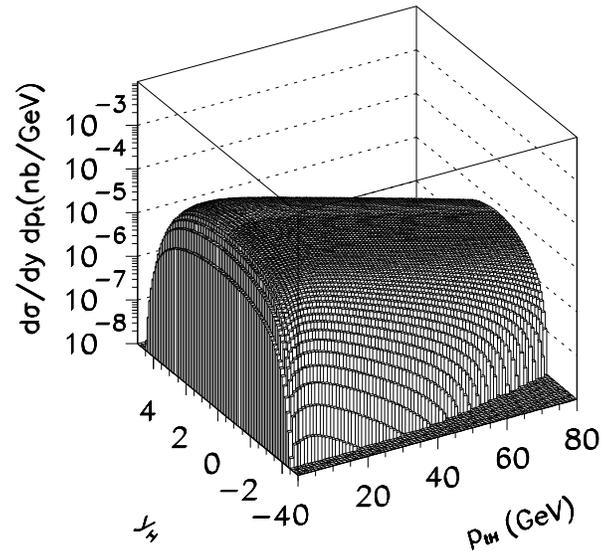}
\caption{\it
A comparison of two-dimensional distributions of Higgs boson for
(a) Kwieci\'nski UGDF, (b) LO b-space resummation.
\label{fig:LHC_y_pt}
}
\end{center}
\end{figure}

It is particularly interesting to compare the results obtained
with the Kwieci\'nski unintegrated gluon distributions
with those obtained from the standard soft-gluon resummation method.
In Fig.\ref{fig:LHC_y_pt} we show two-dimensional distributions in
($y,p_t$).
The distribution obtained with the Kwieci\'nski UGDF decrease
less rapidly with Higgs transverse momentum than the distribution
obtained in the standard soft-gluon resummation. This is partially
due to the different choice of the factorization scale
in both methods.

\begin{figure}[htb] % Figure 10
\begin{center}
    \includegraphics[width=6.5cm]{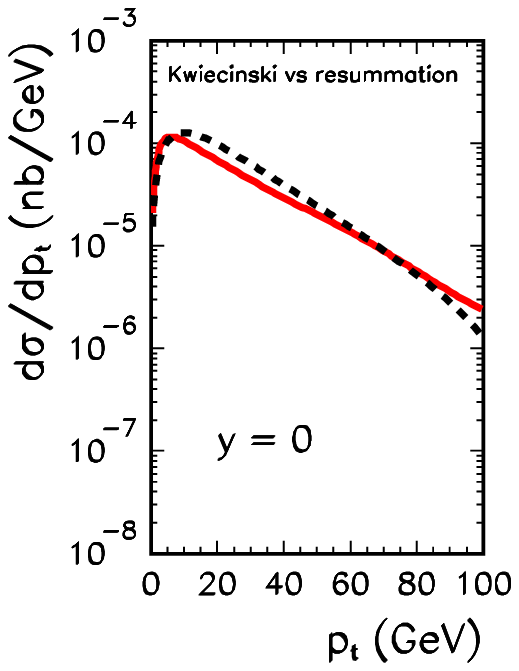}
    \includegraphics[width=6.5cm]{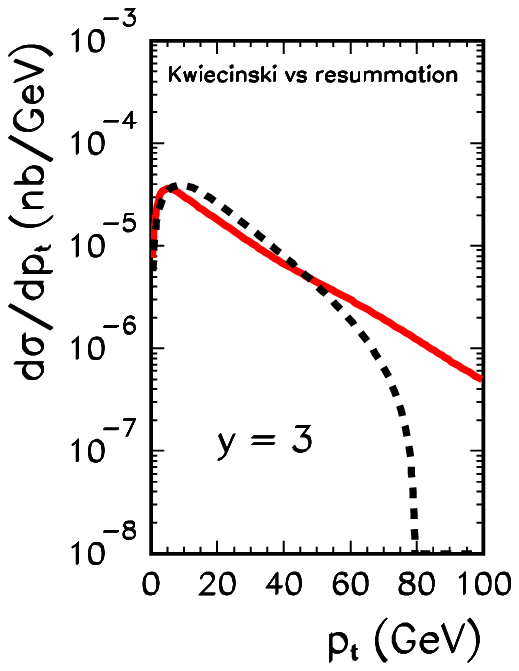}
\caption{\it
A comparison of Higgs transverse momentum distribution
calculated with Kwieci\'nski UGDF (red, solid) and
LO b-space resummation (black, dashed) for different rapidities:
y = 0 (left panel) and y= 3 (right panel). 
\label{fig:kwiecinski_versus_resummation}
}
\end{center}
\end{figure}

A more detailed comparison is made in
Fig.\ref{fig:kwiecinski_versus_resummation} where we have selected two
rapidities $y = 0$ and $y = 3$. While at y = 0 the result obtained
with Kwieci\'nski distributions and that obtained within the standard
resummation method almost coincide, they become quite
different at y = 3 and $p_t >$ 60 GeV. However, the applicability
of both methods at large transverse momenta is not obvious.  
The result for y=3 at large transverse momenta obtained
with Kwieci\'nski UGDF seems more trustworthy than that
obtained within standard resummation method.
The problems of the standard resummation method at large transverse
momenta may be caused by some somewhat arbitrary prescriptions
used as discussed in section 2.

\begin{figure}[htb] % Figure 11
\begin{center}
    \includegraphics[width=6.5cm]{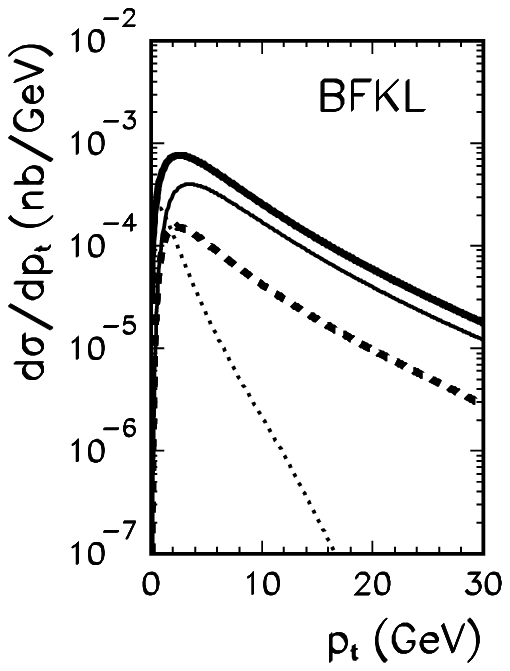}
    \includegraphics[width=6.5cm]{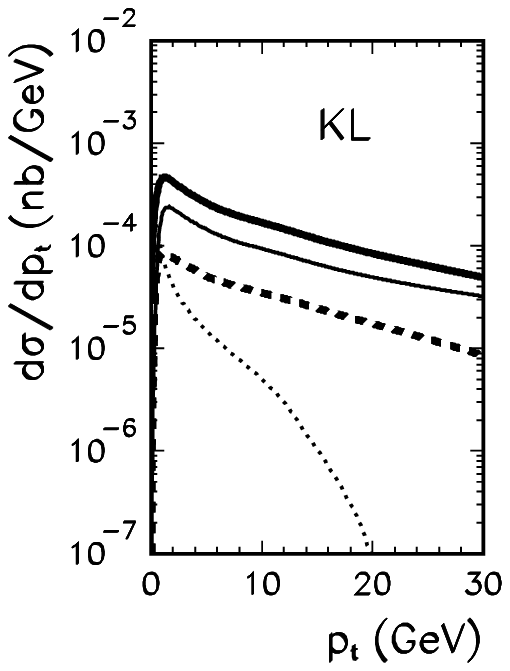}
    \includegraphics[width=6.5cm]{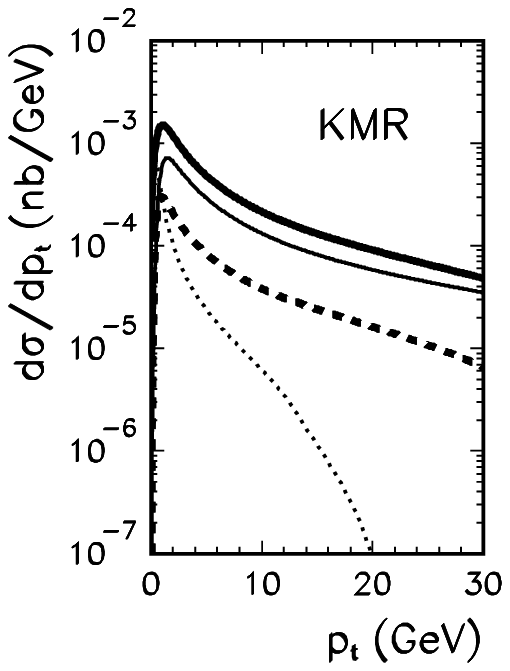}
    \includegraphics[width=6.5cm]{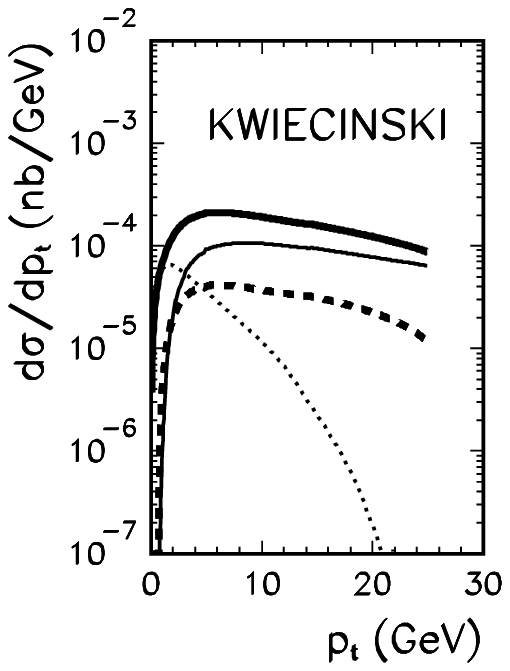}
\caption{\it
Decomposition of the transverse momentum distribution of Higgs boson
at LHC energy W = 14 TeV and -1 $< y <$ 1 into four regions specified
in the text. The thick solid line is a sum of all 4 contributions,
thin solid -- region I, dashed -- region II+III) and dotted --
region IV.
\label{fig:LHC_deco_pt_midrapidity}
}
\end{center}
\end{figure}

\begin{figure}[htb] % Figure 12
\begin{center}
    \includegraphics[width=6.5cm]{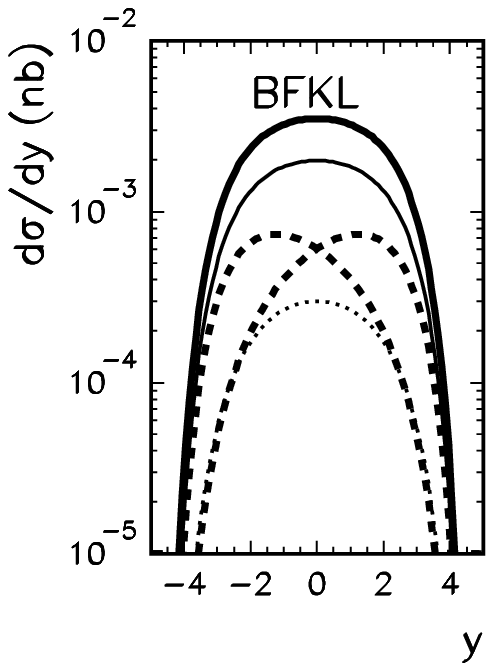}
    \includegraphics[width=6.5cm]{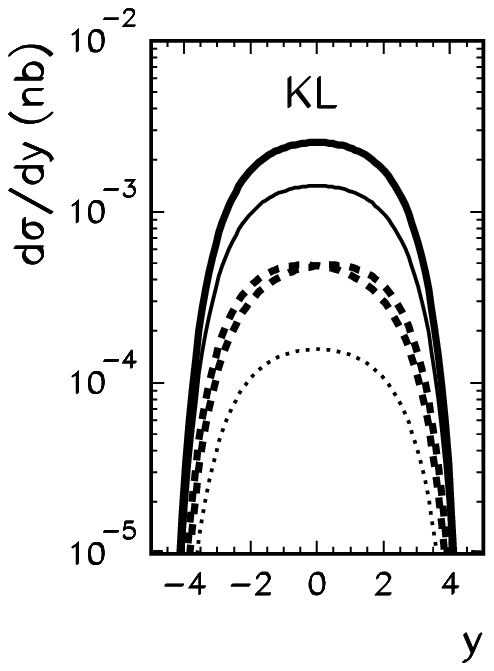}
    \includegraphics[width=6.5cm]{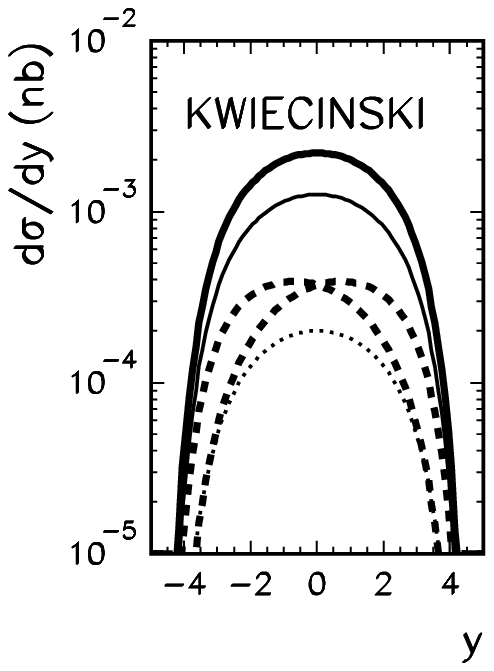}
\caption{\it
Decomposition of the rapidity distribution of Higgs boson
at LHC energy W = 14 TeV and into four regions specified
in the text. In this calculation $p_t <$ 40 GeV.
The thick solid line is a sum of all 4 contributions,
thin solid -- region I, dashed -- region II or III) and dotted --
region IV.
\label{fig:LHC_deco_y}
}
\end{center}
\end{figure}

In order to better emphasize the differences between different UGDF's
we separate the contributions to integral in
Eq.(\ref{momentum_representation_2}) from four 
different disjoint and complementary kinematic regions:
\begin{itemize}
\item (I)   $\kappa_1 < p_t$ and $\kappa_2 < p_t$,
\item (II)  $\kappa_1 < p_t$ and $\kappa_2 > p_t$,
\item (III) $\kappa_1 > p_t$ and $\kappa_2 < p_t$,
\item (IV)  $\kappa_1 > p_t$ and $\kappa_2 > p_t$,
\end{itemize}
where $\kappa_1$ and $\kappa_2$ are transverse momenta of the last
gluons in the ladders and $p_t$ is transverse momentum of the produced
Higgs boson.
In Fig.\ref{fig:LHC_deco_pt_midrapidity} we present the decomposition
of the Higgs cross section $\frac{d \sigma}{d p_t}$ into those four
regions as a function of
Higgs transverse momentum. Here we limit to midrapidities
(-1 $< y <$ 1) only. In the case of Kwieci\'nski UGDF, first
the Fourier transform from the b-space to $\kappa_t$-space was calculated
and the results were stored on the grid in x and $\kappa_t^2$.
The grid was used then for interpolation when using formula
(\ref{momentum_representation_2}) with the extra conditions
on transverse momenta specified above.
It is interesting to note that at larger
transverse momenta the contributions from regions II, III and IV are
completely negligible. The other contributions are important
only at very low transverse momenta.
For completeness in Fig.\ref{fig:LHC_deco_y} we show
similar decomposition as a function of rapidity. In this calculation
we have limited Higgs transverse momenta to $p_t <$ 40 GeV.
The contribution of region I dominates at midrapidities.
The contribution of asymmetric (in $\kappa_1$ and $\kappa_2$) regions
II and III becomes important at very forward or very backward
Higgs production.
The contribution of the region IV is negligible almost
everywhere. The proportions of contributions corresponding to the four
specified above regions differ significantly for different UGDF's.

%--------------------------------------------------------
\subsection{Estimates of higher-order effects}
%--------------------------------------------------------

\begin{figure}[htb] % Figure 13
\begin{center}
    \includegraphics[width=6.0cm]{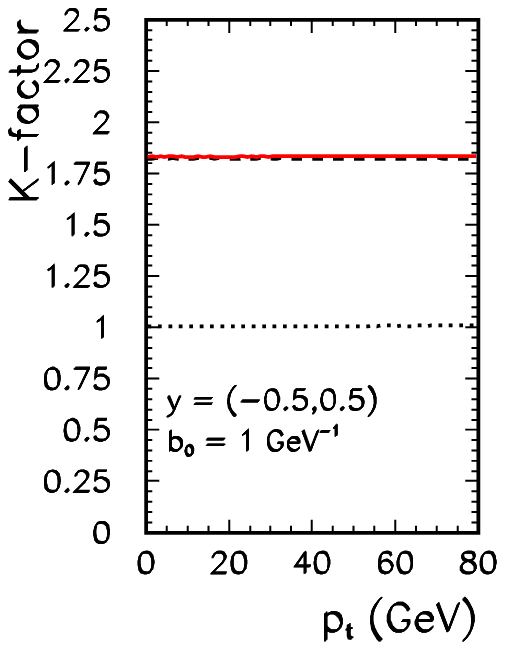}
    \includegraphics[width=6.0cm]{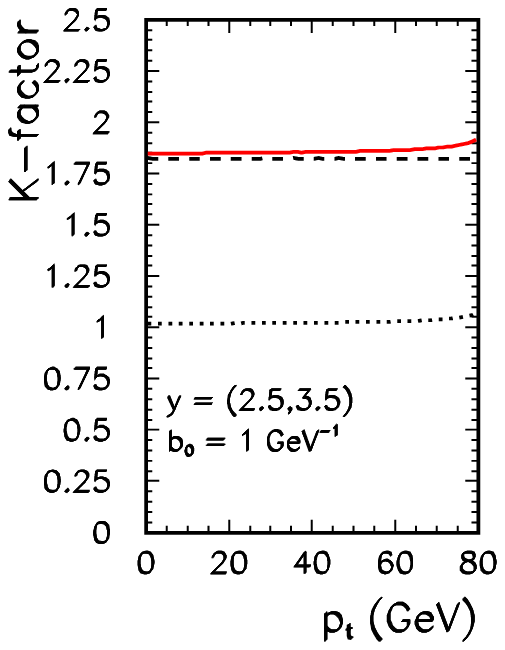}
\caption{\it
NLO K-factor in the soft-gluon-resummation formalism
as a function of Higgs transverse momentum
for two different bins of rapidity. The dashed line
corresponds to separated gluonic contributions, while
the dotted line to separated quarkish contributions.
The solid line is for both effects included simultaneously. 
\label{fig:K-factor_pt}
}
\end{center}
\end{figure}

\begin{figure}[htb] % Figure 14
\begin{center}
    \includegraphics[width=10.0cm]{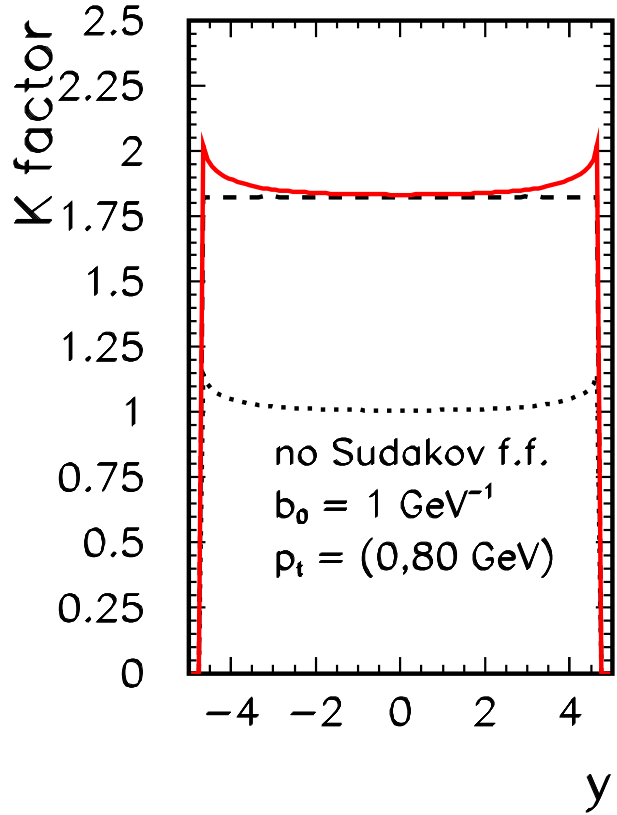}
\caption{\it
NLO K-factor in the soft-gluon-resummation formalism
as a function of Higgs rapidity.
In this calculation $p_t <$ 80 GeV.
The meaning of the curves here is the same as in the previous
figure.
\label{fig:K-factor_y}
}
\end{center}
\end{figure}

In the present paper we have limited to leading-order approach
only. This was dictated by the fact that until now only leading-order
approach was used to describe the HERA data in terms of UGDF's. Furthermore
the consistent next-to-leading order analysis is rather complicated.
We leave the next-to-leading order analysis for a future study.
Instead, we wish to visualize (estimate) the NLO corrections in a
similar soft-gluon-resummation approach where the relevant formalism was
worked out in detail \cite{resummation}.
In Fig.\ref{fig:K-factor_pt} we present the relevant
soft-gluon-resummation K-factor as a function of Higgs transverse
momentum for -0.5 $< y <$ 0.5 (panel a) and 2.5 $< y <$ 3.5 (panel b).
In this calculation the Gaussian form factor (see Eq.(\ref{formfactor}))
with $b_0 =$ 1 GeV$^{-1}$ was used.
The dashed line includes only gluonic NLO corrections to partonic
function, the dotted line exclusively the quark NLO corrections
and the solid line includes all NLO effects altogether.
At midrapidities the gluonic effects are absolutely dominant
and enhance the leading-order cross section by about 80 \%.
The quark corrections are at the level of 1 \%
and can be numerically neglected. They become sizable (of the order of
10 \%) at very forward and very backward rapidities.
In Fig.\ref{fig:K-factor_y} we present a corresponding K-factor
(K = NLO/LO) as a function of Higgs rapidity.
In this calculations $p_{t,H} <$ 80 GeV was assumed.
Summarizing, to a good approximation the NLO
soft-gluon resummation corrections result in multiplying
the LO cross section for Higgs production by a factor of about 1.8.

\begin{figure}[htb] % Figure 15
\begin{center}
    \includegraphics[width=6.5cm]{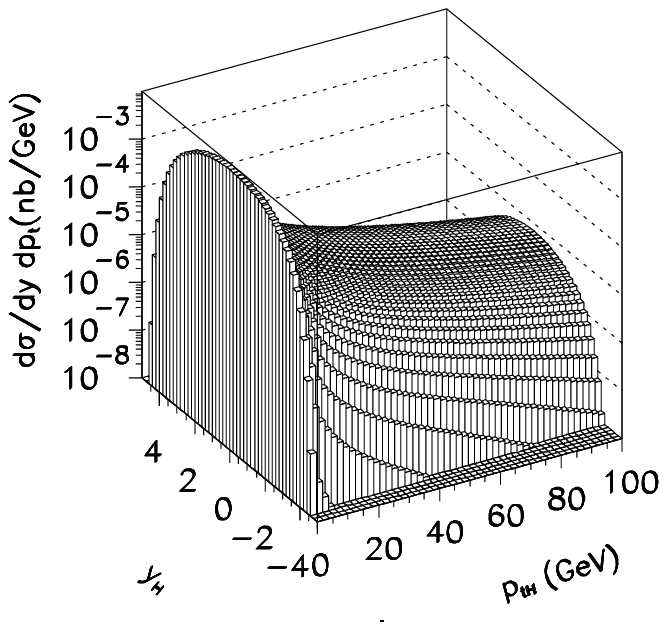}
    \includegraphics[width=6.5cm]{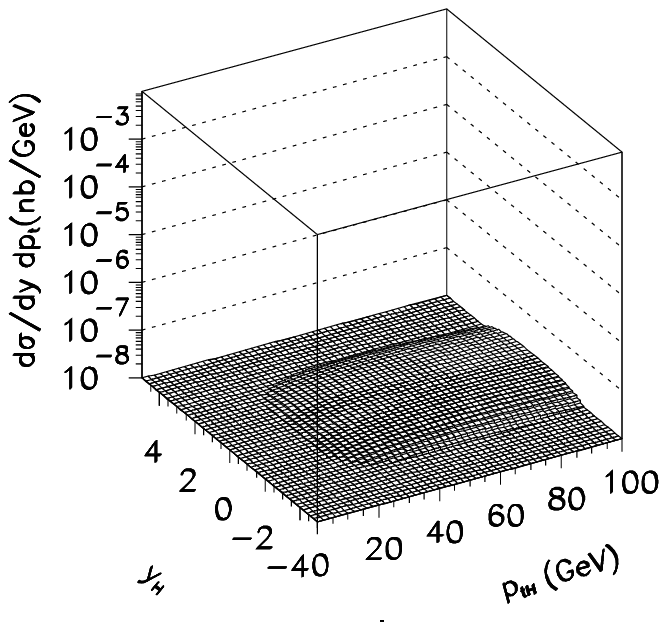}
    \includegraphics[width=6.5cm]{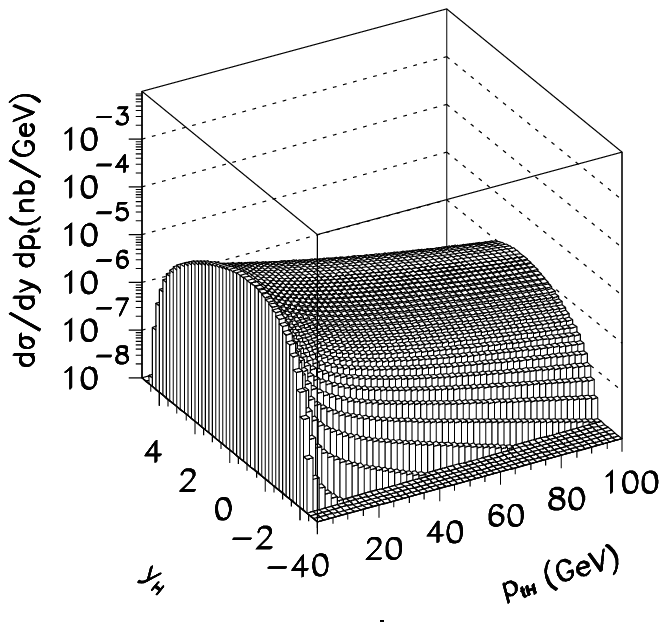}
    \includegraphics[width=6.5cm]{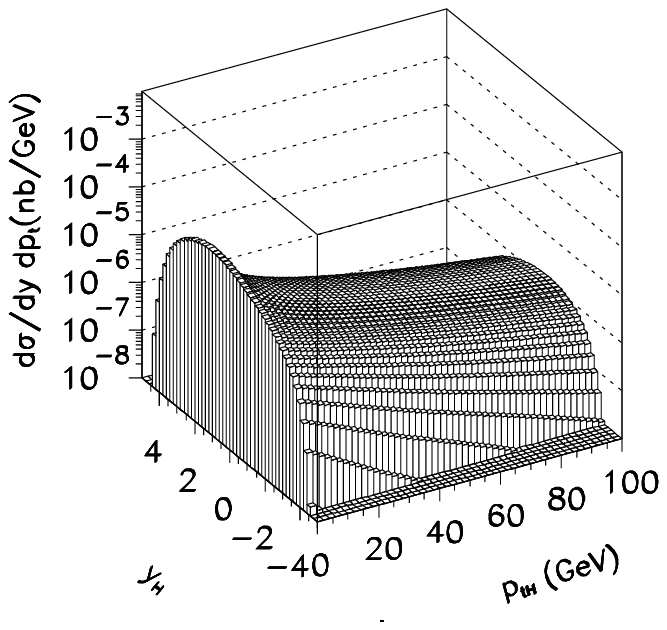}
\caption{\it
Contributions of different subprocesses of the 2 $\to$ 2 type
for W = 14 TeV, respectively for (a) gg, (b) qq, (c) qg and (d) gq.
\label{fig:2to2}
}
\end{center}
\end{figure}

Up to now we have concentrated at relatively small Higgs transverse
momenta. At high transverse momenta the standard 2 $ \to $ 2 mechanisms
take over.
The cross section for perturbative 2 $ \to $ 2 processes with Higgs in
the final state is shown in Fig.\ref{fig:2to2}.
The $gg \to Hg$ dominates at small transverse momenta and
midrapidities. The $qg \to Hq$ and $gq \to Hq$ become comparable
to the first contribution at large transverse momenta and
forward and backward regions, respectively.
The contribution of $qq \to Hg$ is negligible all over
the interesting part of the phase space.

%------------------------------------------------------------------------
\subsection{Weak boson fusion versus gluon-gluon fusion}
%------------------------------------------------------------------------

\begin{figure}[htb] % Figure 16
\begin{center}
    \includegraphics[width=12.0cm]{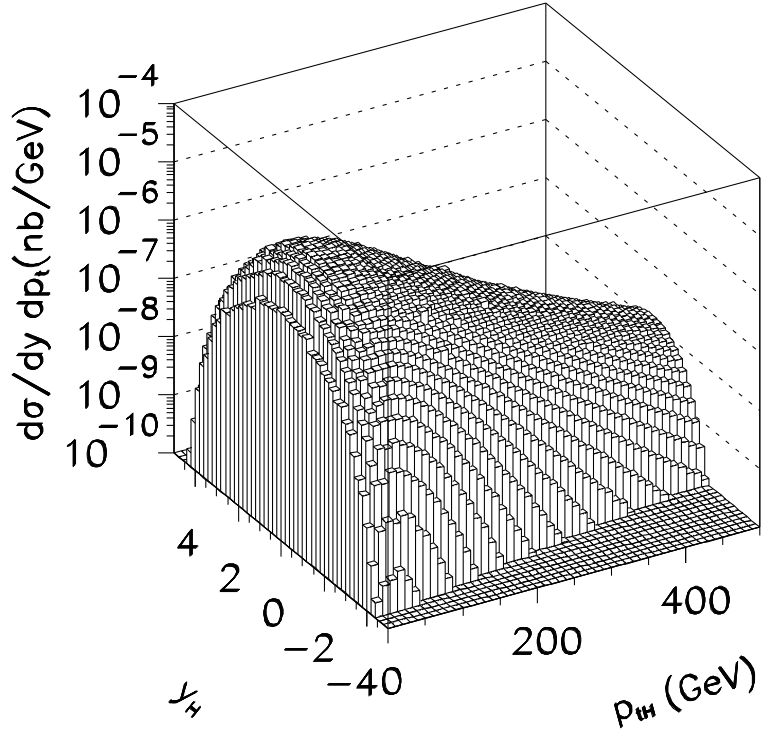}
\caption{\it
Two-dimensional distribution in $y$ and $p_t$ of Higgs from
LO WW fusion process. In this calculation we have taken
$\mu_1^2 = \mu_2^2 = M_H^2$.
\label{fig:WW_ypt}
}
\end{center}
\end{figure}

\begin{figure}[htb] % Figure 17
\begin{center}
    \includegraphics[width=6.5cm]{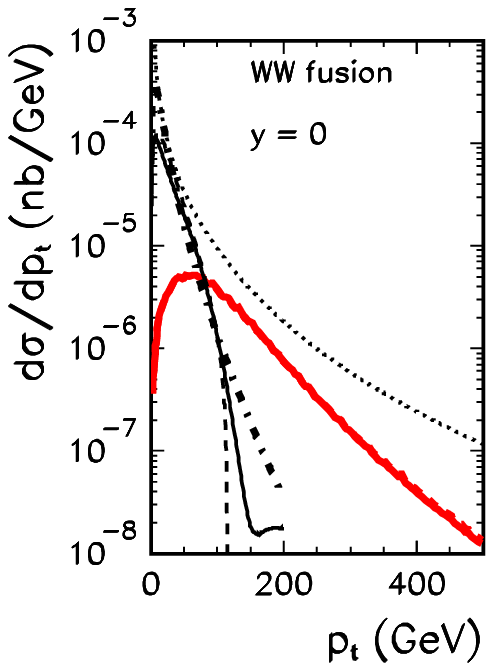}
    \includegraphics[width=6.5cm]{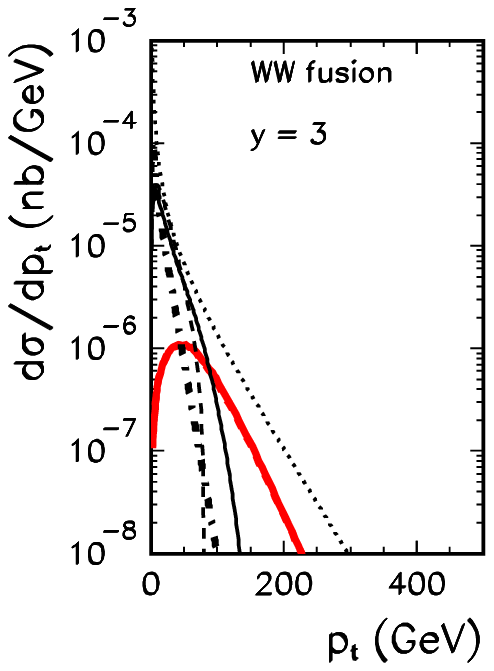}
\caption{\it
Transverse momentum distribution of Higgs boson at LHC energy
W = 14 TeV and y = 0 (left panel) and y = $\pm$ 3 (right panel)
produced in WW fusion (overlaping thick solid and thick dashed
for the two different prescriptions specified in the formalism section),
compared to the corresponding contribution of gluon-gluon fusion:
BFKL (dash-dotted), Kwieci\'nski (thin solid) UGDF,
LO soft-gluon resummation (dashed) and
the perturbative 2 $\to$ 2 collinear contribution (thin dotted).
\label{fig:WW_pt_fixy}
}
\end{center}
\end{figure}

The weak boson fusion is known to be another important ingredient
in total (integrated) cross section for Higgs boson production \cite{book}.
It is interesting what is interrelation between the two dominant
contributions in rapidity and transverse momentum of the Higgs boson.
In Fig.\ref{fig:WW_ypt} we present such a two-dimensional spectrum.
This spectrum is very different from those for the gluon-gluon fusion.
In particular, the maximum of the cross section at
$p_t$ slightly larger than 50 GeV is visible. It is interesting
if the contribution of weak-boson fusion can exceed in some corner
of the phase space the gluon-gluon contribution. In order to quantify
the effect in Fig.\ref{fig:WW_pt_fixy} we present
$\frac{d \sigma}{dy d p_t}$ as a function of Higgs
transverse momentum for y = 0 and y = $\pm$ 3.
The contribution of the WW fusion falls off much faster for
$y = \pm$ 3 than for $y =$ 0. 
The results are almost independent of the choice of the factorization
scale. These results almost coincide (compare thick solid
(prescription (i) above)
and overlaping thick dashed (prescription (ii) above) lines).
For comparison we present a few examples of the gluon-gluon fusion
with BFKL (dash-dotted) and Kwieci\'nski (thin solid)
unintegrated gluon distributions and
standard resummation method with Gaussian form factor and
$b_0$ = 1 GeV$^{-1}$ (dashed).
We find that depending slightly on UGDF and rapidity, above
$p_t \sim$ 50-100 GeV the WW fusion mechanism dominates
over gluon-gluon fusion mechanism. However, the $2 \to 2$ processes
(dotted line) are large, especially at midrapidities.
Only at large rapidities and $p_t >$ 150 GeV the WW fusion
seems to dominate over the other processes. However, there the cross
section is rather small.
Whether this opens a possibility to study the WWH and similarly
ZZH couplings requires further studies.

%--------------------
\section{Conclusions}
%--------------------

In the present paper we have presented predictions for the inclusive
cross section for the Higgs boson production at the LHC energy W = 14
TeV, obtained with the help of different models of unintegrated
gluon distributions used recently in the literature.
Almost all the UGDF's discussed here were obtained based on the analysis
of low-x HERA data for virtual photon -- proton total cross sections.
Although they are almost equivalent in the description of the HERA data,
quite different results have been obtained for Higgs production.
While the structure function data is sensitive to rather low
transverse momenta of the gluon, in the Higgs production, in principle,
one could sample not yet explored region of large transverse momenta
(at large scales). One should remember that the Higgs production
even at large LHC energy is not completely a small-x phenomenon.
The analysis of very forward or very backward Higgs boson production,
in principle, could open a possibility to study UGDF's in a completely
unexplored region of large x. This task, however, is by no means easy
as in this region of phase space both ``small-x'' and ``large-x''
physics is intertangled.

We have shown that for all UGDF's discussed here
the inclusive cross section is dominated
by the configurations with transverse momenta of the gluons smaller
than the transverse momentum of the Higgs
( $\kappa_1 < p_t$ and $\kappa_2 < p_t$ ).

Finally we wish to emphasize that all the potential to study UGDF
in the Higgs production discussed here is at present only conditional
as it implicitly assumes the existence, discovery and good
identification of the Higgs boson in the future experiments at LHC.
We do not need to mention that all this would not be possible if
the Higgsless scenarios (see e.g.\cite{Higgsless_scenarios} and
references therein) turned out to be true.
Even if the Higgs boson is discovered at LHC the statistics
may not be sufficient for precise tests of UGDF. 

\vskip 0.5cm

{\bf Acknowledgements}
We are indebted to Krzysztof Golec-Biernat, El\.zbieta Richter-W\c{a}s
and Tadeusz Szymocha for an interesting
discussion and Krzysztof Kutak for providing us a routine
for calculating unintegrated gluon distributions from Ref.\cite{KuSt04}.
The paper has been improved thanks to suggestions of an anonymous referee.
This work was partially supported by the grant of the Polish
Ministry of Scientific Research and Information Technology
number 1 P03B 028 28.

%-----------------------------------------------------------------------

\newpage

%----------------------------------------------------------------

%----------------------------------------------------------------

%----------------------------------------------------------------

%----------------------------------------------------------------

%----------------------------------------------------------------

%----------------------------------------------------------------

%----------------------------------------------------------------

%----------------------------------------------------------------

%----------------------------------------------------------------

%--------------------------------------------------------------------

%----------------------------------------------------------------

%----------------------------------------------------------------

%----------------------------------------------------------------

%----------------------------------------------------------------

%-----------------------------------------------------------------

%-----------------------------------------------------------------

%----------------------------------------------------------------

%------------------------------------------------------------

%-------------------------------------------------------------

\end{document}